\documentclass[conference]{IEEEtran}
\IEEEoverridecommandlockouts
\usepackage{amssymb}
\usepackage{amsmath,amssymb,amsfonts}
\usepackage{graphicx}
\usepackage{amsmath}  
\usepackage{amssymb}  
\usepackage{geometry}

\newgeometry{left =0.673in,right=0.673in, top= 0.75in,bottom=1.1in}

\ifCLASSINFOpdf
\else
\fi

\begin{document}
\title{Modem Optimization of High-Mobility Scenarios: A Deep-Learning-Inspired Approach}

\author{
    \IEEEauthorblockN{Hengyu Zhang, Xuehan Wang, Jingbo Tan, Jintao Wang}
    \IEEEauthorblockA{Beijing National Research Center for Information Science and Technolog (BNRist), Tsinghua Universityy}
    \IEEEauthorblockA{\{{zhanghen23}, {wang-xh21}\}@mails.tsinghua.edu.cn, \{{tanjingbo},{wangjintao}\}@tsinghua.edu.cn} % 给人名附上邮箱地址
}
\maketitle

\begin{abstract}
The next generation wireless communication networks are required to support high-mobility scenarios, such as reliable data transmission for high-speed railways. Nevertheless, widely utilized multi-carrier modulation, the orthogonal frequency division multiplex (OFDM), cannot deal with the severe Doppler spread brought by high mobility. To address this problem, some new modulation schemes, e.g. orthogonal time frequency space and affine frequency division multiplexing, have been proposed with different design criteria from OFDM, which promote reliability with the cost of extremely high implementation complexity. On the other hand, end-to-end systems achieve excellent gains by exploiting neural networks to replace traditional transmitters and receivers, but have to retrain and update continually with channel varying.  In this paper, we propose the Modem Network (ModNet) to design a novel modem scheme. Compared with end-to-end systems, channels are directly fed into the network and we can directly get a modem scheme through ModNet. Then, the Tri-Phase training strategy is proposed, which mainly utilizes the siamese structure to unify the learned modem scheme without retraining frequently faced up with time-varying channels. Simulation results show the proposed modem scheme outperforms OFDM systems under different high-mobility channel statistics. \footnote{The open source codes are at https://github.com/zhanghy23/ModNet}
\end{abstract}


% \fi
\IEEEpeerreviewmaketitle

\section{Introduction}
In the future 6G-oriented communication systems, multiple high-mobility scenarios, including high-speed train, vehicle-to-vehicle, and satellite communications are envisioned to be further explored \cite{8686339}. The greater dynamics and serious Doppler spread occur when high-speed mobile objectives are involved, bringing a challenge for the design of multi-carrier modulation. Orthogonal frequency division multiplexing (OFDM) is regarded as a vital multi-carrier modulation. It can mitigate the effect of inter-symbol interference (ISI) caused by multi-path delay spread but has difficulty overcoming the inter-carrier interference (ICI) brought by the large Doppler spread \cite{1638663}. Thus, severe performance degradation of OFDM systems occurs in high-mobility scenarios. In order to mitigate the degradation, there have been some modifications of OFDM from different aspects, such as OFDM/offset QAM (OFDM/OQAM) \cite{995073}, lattice-OFDM (LOFDM) \cite{1214833} and filter bank multicarrier (FBMC) \cite{5753092}. These schemes can only alleviate performance degradation to a certain extent, but are unable to achieve excellent performance as they do not change the basic structure of OFDM.

To achieve better communication quality than OFDM, some radical multi-carrier modulation schemes have been proposed. For example, the orthogonal time frequency space (OTFS) modem which exploits full diversity over time and frequency is designed by employing inverse symplectic Fourier transform (ISFFT), i.e. ISFFT-precoded OFDM modulation, outperforming OFDM modulation especially with the large Doppler spread \cite{7925924}. Affine frequency division multiplexing (AFDM) deals with doubly-dispersive channels on the basis of discrete Fourier transform (DAFT) and is also effective for overcoming ICI \cite{9473655}. However, receivers of these modem schemes are extremely complex, such as the minimum mean square error (MMSE) detector for AFDM and message passing (MP) detector for OTFS, requiring more computing time and resources than OFDM. On the other hand, deep learning (DL) presents huge potential in dealing with the complex optimization problem. Recently, there have been considerable works exploiting DL-based methods to transmit symbols. Most of them focus on the end-to-end communication systems, which replace the traditional architecture of transmitters and receivers (including coder/decoder, modulation/demodulation, etc.) with neural networks by directly minimizing the bit error rate \cite{8664650}, \cite{9745781}. However, to adapt to the complex time-varying channels, neural networks deployed at the base station and user sides need to retrain and update frequently. Add to the heavyweight network, it is quite difficult for end-to-end systems to deal with high-mobility scenarios. 

To the best of the authors' knowledge, there has not been a modem scheme that can deal with high-mobility scenarios with similar complexity as OFDM and fixed modulation/demodulation matrices in different channels, which is quite essential for future wireless communication systems. To fill in this gap, we plan to utilize DL methods to optimize a quasi-optimal modem scheme for high-mobility scenarios under a certain design criterion. Similar to OFDM, our proposed modem scheme has the form of unified modulation/demodulation matrices, without changing as channels vary.

In specific, a Modem Network (ModNet) based modem scheme is proposed in this paper. Different from the end-to-end networks, a novel input-output relation, that channel matrices and modulation/demodulation matrices are the inputs and outputs of ModNet respectively, is applied. Thus, the characteristics of channels can be extracted directly and outputs can be applied based on the traditional communication system without neural networks at the transmitters and receivers. Moreover, we propose the Tri-Phase training strategy for ModNet, which helps Modnet output the unified modem schemes while inputting different channels. That is to say, although environments and channels are time-varying, it is not necessary to retrain our ModNet because the unified modem schemes have been designed through our proposed training strategy. As a result, the complexity of our system can be further reduced. In addition, performances of multi-carrier modulation depend on not only the overall transmission quality of all the sub-carriers, but also the worst one of the sub-carriers. Based on that, we design our loss function to directly eliminate the ICI due to Doppler spread. By optimizing Modnet with the loss function, i.e. the optimization criterion, we can gain the quasi-optimal modem scheme under this criterion in theory. Simulation results show that compared to OFDM, the unified modem scheme optimized by Modnet enables the transmission more reliable in high-speed scenarios, and can adapt to the time-varying channel even if some channel statistics change.   

$\textit{Notations:}$ Matrices, vectors, and scalars are denoted by bold uppercase letters, bold lowercase letters, and normal font, respectively. $(\cdot,\cdot)$ following matrices and $(\cdot)$ following vectors indicate the location of the entry in the matrices or vectors. $(\cdot)^H$ stands for Hermitian transpose and $\mathbb{E}[\cdot]$ represents the mathematical expectation. $||\cdot||_F$ denotes the Frobenius norm.
\section{System Model}
We consider the multi-carrier modulation system and use matrix notations. Let $T$ be the total duration of the transmitted signal frame before adding the prefix and the sampling interval is $T/M$. 

We denote $\mathbf{x}\in {\mathbb{C}}^{M\times 1} $ as the symbol vector, where $\mathbb{E}(\mathbf{x})=\mathbf{0}$ and $\mathbb{E}[\mathbf{x}\mathbf{x}^H]=\sigma_s^2\mathbf{I}_M$ hold. In addition, we assume that the modem scheme requires a prefix, which occupies $M_p$ sampling intervals, to combat multi-path propagation like cyclic prefix in OFDM. To simplify the notation, we take $M+M_p=M_L$. The signal after modulation is written as follows: 
\begin{equation}\label{eq1}
{\mathbf{s = \Phi x}},
\end{equation}
where ${\bf{\Phi}} \in {\mathbb{C}}^{M_{L}\times M}$ indicates the modulation matrix.

Denoting by $s(t)$ the transmitted baseband signal after digital-to-analog conversation, the received signal $r(t)$ through the doubly-dispersive channel can be described as follows \cite{7925924}
\begin{equation}\label{eq3}
{r(t)=\iint h(\tau,\nu)s(t-\tau)e^{j2\pi\nu t}d\tau d\nu+w(t)},
\end{equation}
where $h(\tau,\nu)$, $w(t)$ represent the delay-Doppler channel response, and additive noise respectively. It is common that there are a few propagation paths of the channel, so the channel expression is always sparse \cite{8516353}, which can be represented in the form
\begin{equation}\label{eq4}
{h(\tau ,\nu )=\sum_{i=1}^{N_p} h_i \delta (\tau-\tau_i) \delta (\nu-\nu_i) }.
\end{equation}
$\delta(\cdot)$ is the Dirac delta function. The channel $h(\tau ,\nu )$ is composed of $N_p$ propagation paths, where $h_i$, $\tau_{i}$, and $\nu_{i}$ denote the path gain, time delay and Doppler shift of the $i$th path, respectively. Following \cite{995073,1214833,9473655,OTFS_DSE}, we assume the time delay is $l_i\frac{T}{M}$ and $l_i$ is an integer since the sampling interval is sufficiently small. The received signal at the corresponding sampling point can be formulated as a vector $\mathbf{r}=\left \{ r(n) \right \}_{n=-M_p}^{M-1}$
\begin{equation}\label{eq5}
{{r}(n)= \sum_{i=1}^{N_p} h_i{s}(n-l_i)e^{j2\pi nk_i/M}+{w}(n)}.
\end{equation}
where we take $k_i=\nu_iT$ and ${w}(n)\sim\mathcal{CN}(0,\sigma_w^2)$. If ${s}(n-l_i)$ exceeds the range of sampling, we regard it as zero considering the effects of time delay. Further, we rewrite (5) in vector form and describe the channel matrix $\bf{H}$ in detail.
\begin{equation}\label{eq6}
{\bf{r}= \bf{Hs}+\bf{w}},
\end{equation}
\begin{equation}\label{eq7}
{\mathbf{H}= \sum_{i=1}^{N_p}h_i\mathbf{\Delta}^{k_i} \mathbf{\Gamma}_{l_i}},
\end{equation}
where $\mathbf{H},\mathbf{\Delta},\mathbf{\Gamma}_{l_i}\in \mathbb{C}^{M_L\times M_L}$ denote the channel matrix, equivalent delay matrix, and Doppler shifting matrix respectively. $\mathbf{\Delta}$ is a diagonal matrix that brings Doppler shift to the transmitted signal. $\mathbf{\Gamma}_{l_i}$ is composed of zeros and the $M_L-l_i$ order identity matrix to represent the time delay in different paths. 
\begin{equation}\label{eq9}
{\mathbf{\Delta}=\mathrm{diag} [e^{j2\pi\frac{-M_p}{M}},e^{j2\pi\frac{-M_p+1}{M}},...,e^{j2\pi\frac{M-1}{M}}]}
\end{equation}

\begin{equation}\label{eq8}
{
\mathbf{\Gamma}_{l_i} =\begin{bmatrix}
\mathbf{0} & \mathbf{0}\\
\mathbf{I}_{(M_L-l_i)\times(M_L-l_i)} & \mathbf{0}
\end{bmatrix}
}
\end{equation}

\begin{figure*}[!t]
\centering
\includegraphics[width=0.8\linewidth]{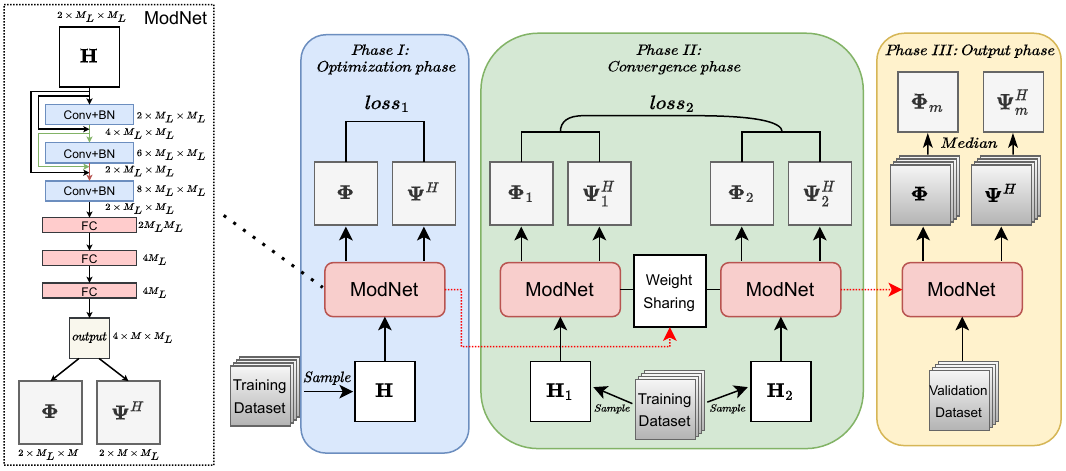}
\caption{The structure of Modem Network (ModNet) and Tri-Phase training strategy. 'Conv+BN' represents a convolution layer followed by batch normalization. 'FC' represents a fully connected layer.}
\label{fig_1}
\end{figure*}

At the receiver, we demodulate the received signal to recover the original symbols. The signal after demodulation is in the form
\begin{equation}\label{eq10}
{\mathbf{y = \Psi}^H \mathbf{ r = \Psi}^H\mathbf{ H \Phi x+ \Psi}^H \mathbf{ w}=\mathbf{H}_e\mathbf{x}+\mathbf{\Psi}^H\mathbf{ w}},
\end{equation}
where ${\mathbf{H}_e=\mathbf{\Psi}^H\mathbf{ H \Phi}}$ denotes the equivalent channel and $\mathbf{\Psi}^H\in \mathbb{C}^{M\times M_L}$ represents the demodulation matrix which removes the prefix of the signal.
\section{Design of ModNet and Training Strategy}
In this section, we introduce our DL methods, including the network structure, loss functions, and the training strategy. Loss functions represent the design criterion of modem schemes so that the network can find a quasi-optimal modem scheme based on that.
% In this section, we ModNet is introduced to solve the optimization problem in this section.We also propose the two-phase training strategy to unify the modem structure fitting different channels.
\subsection{ModNet Structure}
The structure of our designed network called Modem Network (ModNet) is demonstrated on the left of Fig. 1. Differing from end-to-end structures, our goal is to directly design the modem schemes in the matrix form. In specific, channel matrices $\mathbf{H}$ from high-speed scenes are inputs of ModNet, aiming to enable the network to extract related channel information and design suitable modulation/demodulation matrices.

We regard the channel matrix $\mathbf{H}$ as a two-channel picture, whose real part and imaginary part occupy one channel respectively. The input image will pass through three consecutive $7\times 7$ convolution layers, and each convolution operation is followed by the batch normalization. Meanwhile, dense connection is employed among the convolution layers, which can promote the circulation of feature maps and alleviate the over-fitting problem. After that, three fully connected (FC) layers are added to scale the size of the output. Finally, the output image is split into $\mathbf{\Phi}$ and $\mathbf{\Psi}^H$. They also have two channels representing the real part and the imaginary part. Note that each layer except the last FC layer is followed by the activation layer for the non-linearity. Moreover, the energies of $\mathbf{\Phi}$ and $\mathbf{\Psi}^H$ are normalized to $M_L$ and $M$, respectively. Indeed, we focus on proposing this novel input-output relation to reduce the consumption of resources compared to end-to-end systems, and do not claim that the concrete network structure is optimal.

\subsection{Tri-Phase Training Strategy and Loss Functions}
Common training strategies can only let network parameters converge. If inputs are different, outputs tend to vary from each other. Therefore, we cannot get a unified modem scheme using common training strategies and have to retrain and update ModNet frequently due to the time-varying channels in high-speed scenarios, which brings great difficulties in practical application. It is necessary to enable the outputs, i.e. modulation/demodulation matrices, to converge with different inputs. Aiming at this, we propose a Tri-Phase training strategy, including \textit{Optimization phase, Convergence phase, and Output phase.} 
% As Fig.1 shows, phase I is responsible for improving the performance of the learned modem structure, while the work of making the modulation/demodulation matrices tend to be fixed is arranged in phase II.
\subsubsection{\textit{\textbf{Phase I: Optimization phase}}}
Phase I is similar to common training strategies, sampling from the training dataset and updating the network parameters based on the gradient of the loss function. The loss function represents our design criterion and guides ModNet to design a quasi-optimal modem scheme under it. Next, we will introduce the design of the loss function.
\begin{figure*}[!t]
\centering
\includegraphics[width=0.917\linewidth]{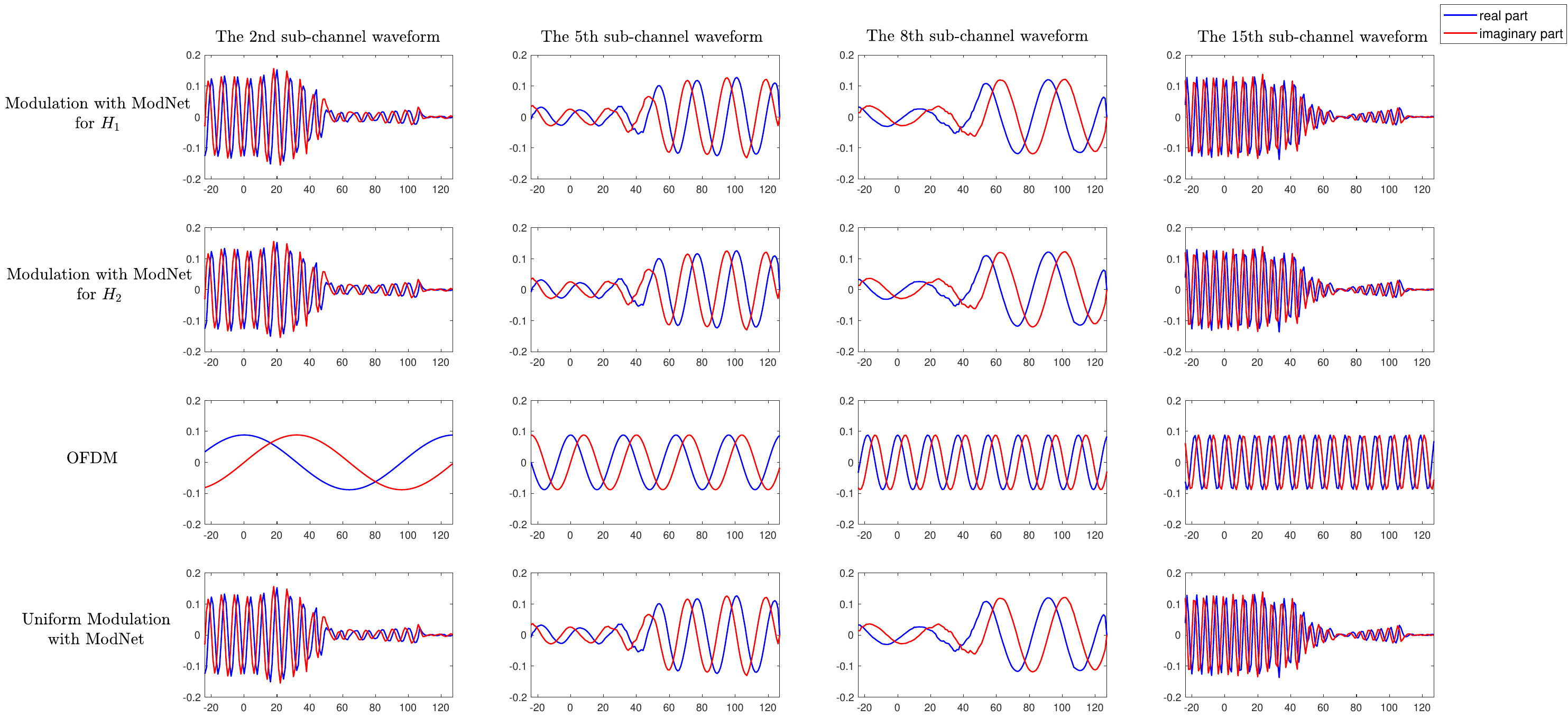}
\caption{Comparisons of the sub-channel modulation waveforms with $v=\mathrm{360km/h}$ and $N_p=4$ between our proposed ModNet and OFDM.}
\label{fig_3}
\end{figure*}
Derived from (9), the received symbols $\mathbf{y}$ can be regarded as the transmission in $M$ equivalent sub-channels. The received symbol from the $m$th equivalent sub-channel is written as follows:  
% \begin{footnotesize}
\begin{equation}
\begin{aligned}
\mathbf{y}(m)&=\mathbf{H}_e(m,m)\mathbf{x}(m)+\sum_{n\ne m}\mathbf{H}_e(m,n)\mathbf{x}(n)\\
&+\sum_{n=-M_p}^{M-1}\mathbf{\Psi}^H(m,n)\mathbf{w}(n),
\end{aligned}
\end{equation}
% \end{footnotesize}

where two sum items represent interference and noise, respectively. Severe Doppler spread aggravates the interference between sub-channels. Using the most straightforward idea, we want to reduce overall system interference and ensure each sub-channel a good communication quality, because not only the general state of the overall system but also the sub-channel with the worst situation can greatly affect the transmission error rate. Based on that, the loss function is taken in the form
\begin{equation}
loss_1 = -[\sum_{m=0}^{M-1}r_{m}(\mathbf{H}_{e})+M*\min_{m}r_{m}(\mathbf{H}_{e})],
\end{equation}
where $r_m(\mathbf{H}_e)$ represents the rate of $m$th equivalent sub-channel and is in the specific expression
\begin{scriptsize}
\begin{equation}   
r_{m}(\mathbf{H}_e)= \log_{2}{(1+\frac{\left |  \mathbf{H}_{e}(m,m)  \right |  ^{2} }{\sum_{n\ne m}\left | \mathbf{H }_{e}(m,n)  \right |  ^{2}+\frac{\sigma_{w}^{2}}{\sigma_{s}^{2}}\sum_{n=-M_p}^{M-1} \left |  \mathbf{\Psi}^H(m,n)   \right |  ^{2}   })}
\end{equation}
\end{scriptsize}
where $\frac{\sigma_{w}^{2}}{\sigma_{s}^{2}}$ represents the reciprocal of signal-to-noise ratio (SNR). In (11), the sum item and min item are applied to guarantee the performance of the overall system and the worst channel, respectively. We add the minus sign in the loss function so that the direction of the gradient descent coincides with the direction of the ICI reduction.

It is worth noting that different design criteria can be taken under different requirements. Here the interference between sub-channels is in our first position. Our approach is compatible with different criteria, that is to say, no matter what the criteria are, they all can be optimized using our proposed DL strategies to achieve a quasi-optimal solution. 

% we directly exploit the ModNet to optimize the modem structure. The channel samples from the training dataset are utilized to update the weights of the ModNet. Based on the OESCR criterion, the loss function is designed as follows:
% % \begin{equation}
% % {loss_1 = f(\mathbf{H}_{e,\mathrm{OFDM}})-f(\mathbf{H}_e)},
% % \end{equation}
% % \begin{equation}
% % {\mathbf{H}_{e,\mathrm{OFDM}}=\mathbf{\Psi}^H _{\mathrm{OFDM}}\mathbf{H} \mathbf{\Phi} _{\mathrm{OFDM}}},
% % \end{equation}
% % where $\mathbf{H}_{e,\mathrm{OFDM}}$ represents the equivalent channel in the OFDM system. 

% In this form of (17), we can clearly judge whether the optimized modulation system performance is better than the OFDM system from the sign of loss function. Our objective is to make the value of the loss function much less than zero. That means the ModNet can find the corresponding modulation/demodulation matrices for the input channel that are superior to OFDM. 

\subsubsection{\textit{\textbf{Phase II: Convergence phase}}}
After phase I, the modulation/demodulation matrices learned by ModNet vary with the input channels. In phase II, we utilize the siamese architecture to obtain a unified modem. 

The siamese architecture is often used in contrastive learning. In specific, the siamese architecture represents that two same networks are used to make a classification. If two samples from different categories are fed into two networks, the high-dimensional distance between the outputs of the two networks is required to be as large as possible. Conversely, if both samples are from the same category, the outputs of the two networks should be as similar as possible.

Different from the original siamese architecture, we simply treat all channel matrices from a channel model as a category, making the outputs of two networks as similar as possible in each sample. In this way, the modem scheme learned by ModNet can adapt to time-varying channels without frequent retraining. In each sample, two different channel matrices $\{\mathbf{H}_1,\mathbf{H}_2\}$ randomly selected from the training dataset are fed into two networks respectively at a time. $\{\mathbf{\Phi}_1,\mathbf{\Psi}_1^H\}$ and $\{\mathbf{\Phi}_2,\mathbf{\Psi}_2^H\}$ correspond to the results of two networks.

During phase II, two networks share the same weight parameters all along. The parameters are updated according to the following loss function.
\begin{equation}
\begin{aligned}
loss_2 &= \alpha[-\sum_{m=0}^{M-1}r_{m}(\mathbf{H}_{e})-M*\min_{m}r_{m}(\mathbf{H}_{e})]\\
&+(1-\alpha)[\left \| \mathbf{\Phi}_1-\mathbf{\Phi}_2 \right \| _{F}^{2}+\left \| \mathbf{\Psi}_1^H-\mathbf{\Psi}_2^H \right \| _{F}^{2}],
\end{aligned}
\end{equation}
where $\alpha$ is a hyper-parameter. The loss function in phase II is composed of two parts. One part is like $loss_1$, which continues to keep the system performance. On the other hand, another part is introduced to shorten the distance between the modem schemes learned from two different channel samples. 

The parameters of the network from phase I are the initial weight parameters in phase II to speed up and help with the training of ModNet. By modifying $\alpha$ to an appropriate value, ModNet can output similar modem schemes in time-varying channels without increasing the interference between sub-channels too much.
% $
% In phase II, two different channel matrices making up a sample $\{\mathbf{H}_1,\mathbf{H}_2\}$ are entered into two ModNets respectively at a time. $\{\mathbf{\Phi}_1,\mathbf{\Psi}_1^H\}$ and $\{\mathbf{\Phi}_2,\mathbf{\Psi}_2^H\}$ correspond to the results of two channels.

% Considering the goal to converge the modulation and demodulation matrices, the loss function is designed as (19).
% \begin{equation}
% \begin{aligned}
% \quad \quad loss_2 &= \alpha [f(\mathbf{H}_{e,\mathrm{OFDM}})-f(\mathbf{H}_e)]\\ 
% &+ (1-\alpha)[g(\mathbf{\Phi}_1 , \mathbf{\Phi}_2)+g(\mathbf{\Psi}_1^H ,\mathbf{\Psi}_2^H)],
% \end{aligned}
% \end{equation}
% \begin{equation}
% {g(\mathbf{X}_1,\mathbf{X}_2)=\left \| \mathbf{X}_1-\mathbf{X}_2 \right \| _{F}^{2} }
% \end{equation}

% Compared to $loss_1$, the function $loss_2$ is composed of two parts and $\alpha$ is a hyper-parameter balancing them. The first part on the right of equation (19) is similar to $loss_1$, which still keeps the system performance. Besides, the function $g(\cdot, \star)$ is to evaluate the gap between two modem structures learned from different channels. Therefore, the second part on the right of equation (19) can force the modulation/demodulation matrices to further converge to the unified form.

% With the help of the two-phase training strategy, our ModNet can find the quasi-optimal modem structure easily. After the convergence of the network, the validation data samples are put into the ModNet. Then we select the median of all the outputs and normalize it to be the final modulation/demodulation matrices.

\subsubsection{\textit{\textbf{Phase III: Output phase}}}
Different from the first two phases, the model parameters are not updated in phase III. We directly adopt the model parameters after the convergence of phase II. Besides, the samples from the validation dataset instead of the training dataset are fed into Modnet to make the designed modem schemes more general. Due to the effect of the siamese structure in phase II, the modulation/demodulation matrices learned by the samples in the validation set are also very close. We take the median of each element in these modulation/demodulation matrices, and the energies of $\mathbf{\Phi}$ and $\mathbf{\Psi}^H$ are normalized to $M_L$ and $M$, respectively. Finally, we obtain only one modem scheme that can adapt to different channels.
\begin{figure}[!t]
\centering
\includegraphics[width=0.765\linewidth]{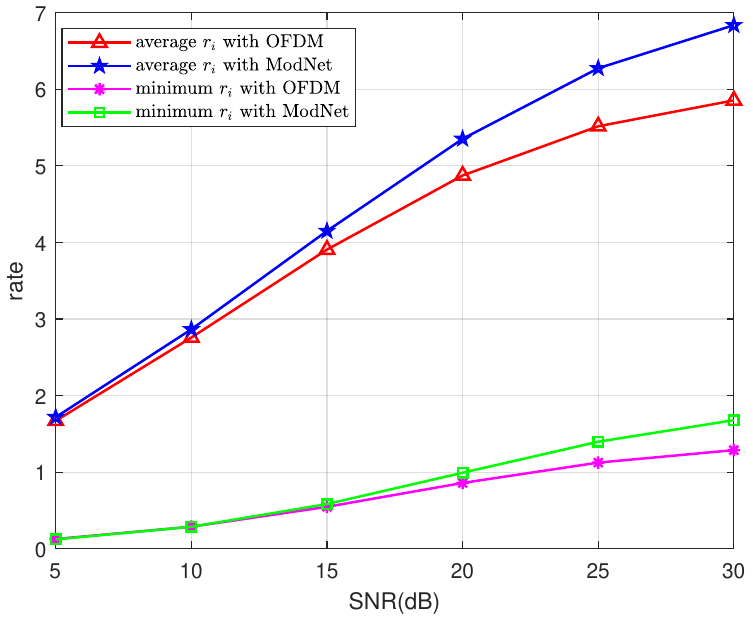}
\caption{The average and minimum equivalent sub-channel rate $r_i$ with $v=\mathrm{360km/h}$ and $N_p=4$.}
\label{fig_3}
\end{figure}
\section{Simulation Results and Analysis}
\subsection{Experiment Settings}
The parameters of channel models and our proposed ModNet are discussed in this section. We present the channel statistics in Table I, most of which follow the typical values in \cite{OTFS_DSE}. The complex gain of each path follows the distribution of $\beta_i\sim \mathcal{CN}(0,1/N_p) $ and is randomly generated. Maximum delay grid means that the delay grid of each path is randomly selected from integers between 0 to $l_{\max}$, which is usually easy to estimate. Besides, we set $M_L=152$ and $M_p=24$ in our methods and OFDM for a fair comparison.

The training dataset and validation dataset include 10,000 and 2,000 samples respectively, generated at the relative speed of 360km/h and 4 propagation paths in the channel. To further demonstrate the advantages of our approach, we generate another 20,000 channel matrices as the testing dataset to present the following numerical results, which are unknown to the learned modem scheme before.

Adam optimizer with settings $\{\beta_1=0.9,\beta_2=0.999,lr=1e-3,\epsilon= 1e-8\}$ is applied to train ModNet. Moreover, $\{epochs=500, batch size=200\}$ are set in both phase I and II. In phase I, the training and validation dataset can be directly used. We need to randomly pair each channel matrix with another different one so that they make up a new sample for training in phase II. ${\alpha=0.005}$ is set in $loss_2$.

\subsection{Analysis of The Uniform Modem Scheme}
In order to better explain the help of ModNet for modem optimization, we randomly select waveforms of 4 sub-channels which are shown in Fig. 2.

The pictures of the first two rows show the modulation mode learned by our proposed ModNet in phase II for two different channels $\mathbf{H}_1$ and $\mathbf{H}_2$. The time delays and Doppler shifts of 4 propagation paths in $\mathbf{H}_1$ are $[5.21,4.69,4.69,2.08]\mathrm{\mu s}$ and $[-1297,-460,-386,-1329]\mathrm{Hz}$, while those in $\mathbf{H}_2$ are $[3.64,2.60,1.56,1.04]\mathrm{\mu s}$ and $[1330,-803,-1185,1301]\mathrm{Hz}$. In phase III, we select the normalized median of outputs with 2,000 validation channels as the final modem structure, whose modulation waveforms are presented in the last row. Comparing the waveforms of the first, second, and fourth rows, we find that they are almost the same. That is to say, the siamese structure and $loss_2$ function in the training phase II have a huge contribution to narrowing the gap between modem structures learned from different channels (The results of the demodulation waveforms are also similar, which are omitted due to page limitation).

\begin{figure}[!t]
\centering
\includegraphics[width=0.8\linewidth]{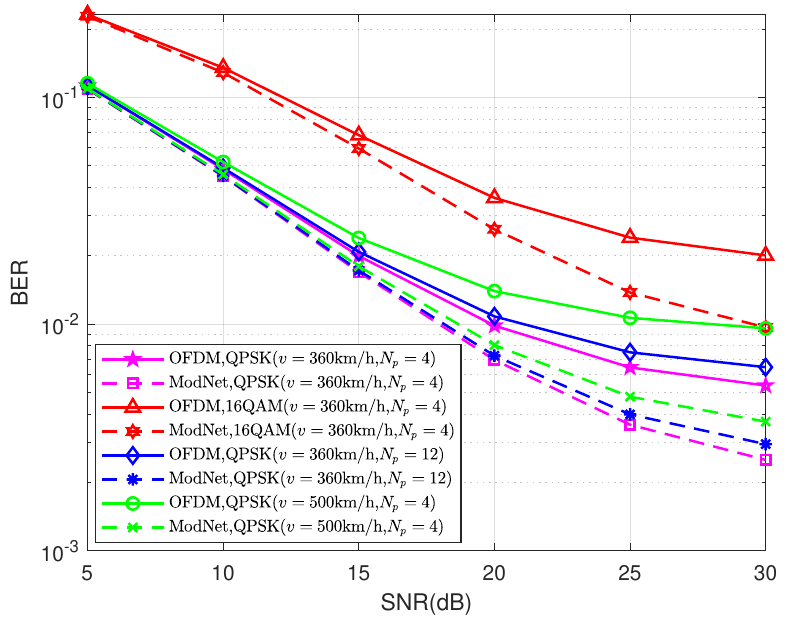}
\caption{BERs of OFDM and modem with ModNet in different alphabets (QPSK and 16QAM) and $v=360,500\mathrm{km/h}$, $N_p=4,12$ .}
\label{fig_3}
\end{figure}
\begin{table}[]
\caption{Channel Parameters\label{table1}}
\renewcommand{\arraystretch}{1.1}
\centering
\setlength{\tabcolsep}{5mm}{
\begin{tabular}{ll}
\hline
Parameter             & Value      \\ \hline
Carrier frequency ($f_c$)     & 4GHz       \\
Subcarrier spacing ($\Delta f$)   & 15kHz      \\
Number of subcarriers ($M$) & 128        \\
UE speed ($v$, $\mathrm{km/h}$)       & 360,500    \\
Modulation alphabet   & QPSK,16QAM \\
Maximum time delay grid ($l_{\max}$)   & 20         \\
Number of paths ($N_p$)       & 4,12          \\ \hline
\end{tabular}}
\end{table}
On the third row, we show the OFDM waveform of 4 sub-carriers. For time-dispersive systems and frequency-dispersive systems, the orthogonal eigenfunctions are complex sinusoids with cyclic prefix and impulse signals with infinite frequency extension, respectively \cite{oppenheim1997signals}. Sub-carriers in OFDM are orthogonal complex sinusoids that only consider the time dispersion and ignore the frequency dispersion. As a result, OFDM has difficulty dealing with double-dispersive systems. By comparison, modulation waveforms with ModNet have a trend to only take up part of the time domain. It tries to make a balance between complex sinusoids and impulse signals, which serve as better sub-carriers for doubly-dispersive channels.   

\subsection{Superiority of Our Methods}
Through simulation based on the testing dataset, we first compare the average and minimum equivalent sub-channel rates between OFDM and our proposed modem scheme in Fig.3. It is clear that the modem scheme learned by ModNet contributes to the higher average and minimum rate. With the help of ModNet, it can improve approximately 20\% of the average equivalent sub-channel rate at each SNR than OFDM. Meanwhile, it also guarantees the quality of the worst sub-channel, which is similar to OFDM. To be added, all the simulation results are based on the training at $\mathrm{SNR=20dB}$, which indicates that our optimized modem scheme has good robustness.       

Further, the performance of the unified modem scheme learned by ModNet is assessed by presenting BER compared with OFDM. Assuming the perfect channel state information (CSI) available at the receiver, the low-complexity LMMSE-based equalization technique is applied as
\begin{equation}
{\hat{\mathbf{x}}(m)=\frac{\mathbf{H}^{*}_e(m,m)}{\left |\mathbf{H}_e(m,m) \right |^2 +\frac{\sigma_{w}^{2}}{\sigma_{s}^{2}}*\sum_{n=-M_p}^{M-1} \left |  \mathbf{\Psi}^H(m,n)   \right |  ^{2} }\mathbf{y}(m)},
\end{equation}
where $\hat{\mathbf{x}}(m)$ denote the $m$th symbol after equalization.

Our modem scheme is learned with the channel parameters of $l_{max}=20$, $v=360\mathrm{km/h}$, and $N_p=4$. Indeed, since $l_{max}$ is only a maximum limit, we can set it as large as possible during training to enable Modnet to learn a modem scheme that can be applied to channels with different time delays. However, $v$ and $N_p$ are factors that can influence the channel statistics and change frequently. As a result, we should consider whether our modem scheme can adapt to the change of channel statistics.  

In Fig. 4, the BER of OFDM and modem with ModNet in different channel environments are explored. Our proposed modem has a lower BER under all the SNRs when the testing channel statistics are the same as the training ones. Especially under $\mathrm{SNR=30dB}$, our methods reduce BER by approximately 50\%, reaching $2.5\times10^{-3}$. Moreover, when the channel statistics change, such as $v$ changes from 360$\mathrm{km/h}$ to 500$\mathrm{km/h}$, and $N_p$ changes from 4 to 12, our modem scheme still shows better performance than OFDM. This means that our modem scheme can not only deal with time-varying channel environment, but also has good generalization under different channel statistics. 

% \section{Discussion}
% In this section, we discuss the potential of our work and point out that some malleable works deserve conducting. 

% We design the OESCR criterion, guiding us to find the corresponding modem structure. Indeed, there are many other criteria that can be designed for different goals. No matter what the criteria are, they all can be optimized using our proposed DL strategies to achieve a quasi-optimal solution. Our work provides a new paradigm for modem optimization.

% Besides, we focus on proposing this new paradigm for a communication problem, instead of designing the complex network structures. We believe that if a more proper network model is applied to this optimal problem, there will be a better performance. That is a potential direction for future works.
\section{Conclusion}
To address the performance degradation of OFDM under high-mobility channels, we proposed the ModNet and Tri-Phase training strategy, which can optimize a modem scheme without replacing the traditional transmitters and receivers of neural networks. We also exploit the potential of siamase structure and design the loss function to unify the modem scheme adapting to time-varying channels. Similar to OFDM, our modem scheme has a fixed form of modulation/demodulation matrices even if the channels are different. Simulation results proved that our modem schemes outperform OFDM systems in different channel environments and presented good robustness. Our future work will focus on the further optimization of modem schemes and consider the design of channel estimation, so that the proposed modem schemes can better meet the needs of practical scenarios.


%\bibliographystyle{IEEEtran}

\bibliographystyle{IEEEtran}
\bibliography{IEEEabrv,ref}

\end{document}